\documentclass[final,5p,times,twocolumn]{elsarticle}

\usepackage{rotating}
\usepackage{amssymb}
\usepackage{amsmath}
\usepackage{amsthm}
\usepackage{array}
\usepackage{float}
\usepackage{rotating}
\usepackage{subcaption} 
\usepackage{hyperref}
\begin{document}

\begin{frontmatter}



\title{Hierarchical Bayesian Modeling of Dengue in Recife, Brazil (2015–2024): The Role of Spatial Granularity and Data Quality for Epidemiological Risk Mapping}


\author{Marcílio Ferreira dos Santos} 

\affiliation{organization={UFPE},
            addressline={}, 
            city={Caruaru},
            postcode={}, 
            state={PE},
            country={Brazil}}

\author{Andreza dos Santos Rodrigues de Melo}
\affiliation{organization={UFPE},
            addressline={}, 
            city={Recife},
            postcode={}, 
            state={PE},
            country={Brazil}}

\begin{abstract}
Dengue remains one of Brazil’s major epidemiological challenges, characterized by pronounced intra-urban inequalities and a strong influence of climatic and socio-environmental factors. This study analyzed confirmed dengue case data in Recife from 2015 to 2024 (a ten-year period), applying a Bayesian hierarchical spatio-temporal model via \texttt{R-INLA}, with a BYM2 spatial structure and an RW1 temporal component. Covariates included population density, average number of residents per household, proportion of drainage channels, relative income, lagged precipitation, and mean temperature. Results indicated a positive effect of population density and household size on dengue risk, and a negative effect of income and channel presence. One-month lagged precipitation showed a positive association, whereas mean temperature exhibited an inverse relationship, suggesting thermal thresholds for vector inhibition. The model showed good fit (\textbf{DIC = 65,817}; \textbf{WAIC = 64,506}) and stable convergence, with moderate residual spatial autocorrelation ($\Phi = 0.06$) and a smooth temporal trend between 2016 and 2019. The spatio-temporal estimates revealed persistent high-risk clusters concentrated in the northern and western zones of Recife, overlapping with areas of higher population density and social vulnerability. Beyond reproducing historical patterns, the Bayesian hierarchical model enables \textbf{probabilistic projections} for subsequent years, providing a scientific basis for \textbf{early warning systems and territorial health planning}. Compared with classical models (GLM, SAR, GWR, GTWR), \texttt{INLA} offers key advantages: (i) explicit integration of parameter and latent effect uncertainty; (ii) simultaneous incorporation of spatial and temporal dependence; and (iii) full inference and forecasting capabilities with credible intervals. These features make the hierarchical Bayesian approach a powerful tool for urban epidemiological surveillance, enhancing the identification of critical areas and supporting evidence-based decision-making in public health management. 
\end{abstract}

\begin{graphicalabstract}
\includegraphics[width=\textwidth]{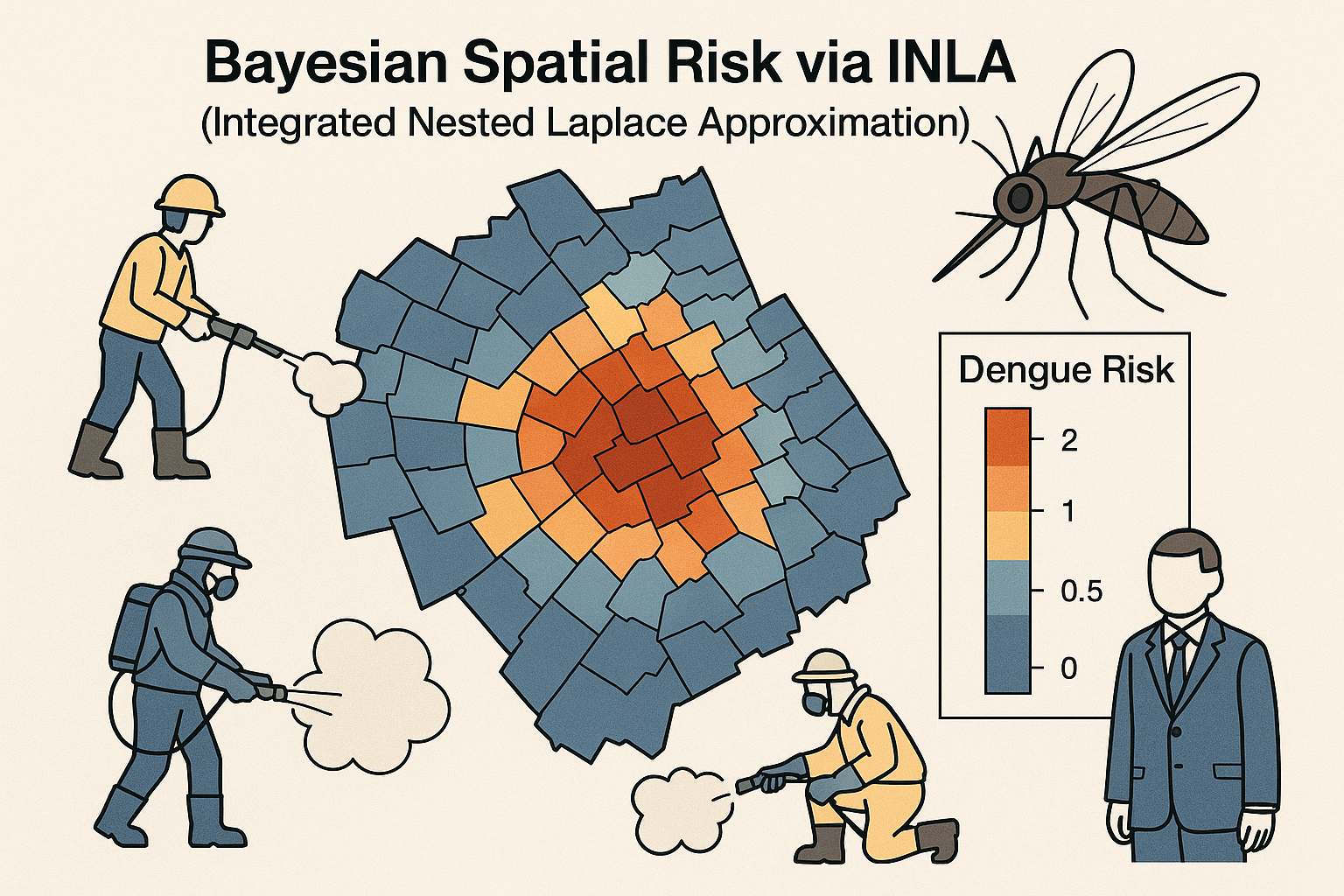}
\end{graphicalabstract}

\begin{highlights}
\item Developed a Bayesian hierarchical space--time model using the \texttt{R-INLA} framework to estimate dengue risk in Recife (2015--2024). 
\item Integrated census, socioeconomic, and climatic variables (population density, household size, income, canal proportion, temperature, and lagged rainfall). 
\item Spatial structure modeled with the BYM2 specification and temporal evolution with a first-order random walk (RW1). 
\item Found positive associations between dengue incidence and population density, household size, and rainfall lag, and negative effects of income and temperature. 
\item The model achieved strong goodness-of-fit (\textbf{DIC = 65,817}; \textbf{WAIC = 64,506}) with stable convergence and moderate residual spatial dependence ($\Phi = 0.06$). 
\item Highlighted persistent high-risk clusters in the North and West regions of Recife, consistent with known vulnerability zones. 
\item Demonstrated that Bayesian hierarchical models outperform classical spatial econometric approaches by providing full uncertainty quantification, spatial–temporal smoothing, and probabilistic forecasting for early-warning systems. 
\end{highlights}


\end{frontmatter}



\section{Introduction}

Dengue remains one of the major public health challenges in tropical and subtropical regions, showing significant expansion in densely populated urban centers. Its spatial dynamics are complex, resulting from the interaction of environmental, socioeconomic, and structural factors—such as precipitation, vegetation cover, sanitation, and housing density—that influence the proliferation of Aedes aegypti. Understanding the territorial distribution of the disease is therefore essential for planning preventive actions and efficiently allocating health resources \cite{barcellos2001}.

Advances in spatial and spatio-temporal modeling have enabled the integration of geographic, climatic, and social data to estimate infection risk across multiple scales. Classical models such as the Spatial Autoregressive Model (SAR) and the Conditional Autoregressive Model (CAR) incorporate spatial dependence in epidemiological contexts but exhibit limitations in flexibility, uncertainty treatment, and the incorporation of hierarchical effects. In contrast, modern Bayesian approaches—particularly the Integrated Nested Laplace Approximation (\texttt{R-INLA})—provide efficient and accurate inference for latent Gaussian models, allowing the representation of complex spatio-temporal structures with both statistical rigor and reduced computational cost \cite{rue2009, blangiardo2015, lindgren2011}.

In Brazil, recent applications of INLA to vector-borne diseases have contributed to identifying critical areas and analyzing socio-environmental determinants \cite{honorio2009, oliveira2022}. However, most of these studies rely solely on raw counts or population-based rates, without explicitly adjusting for territorial exposure. This limitation can introduce bias in large spatial units, where risk is more dependent on environmental structure than on population density.

Thus, a gap remains in the literature regarding the explicit incorporation of \textbf{spatial granularity} and \textbf{territorial exposure} in Bayesian modeling. Few studies treat risk as an adjusted environmental density—that is, the ratio between observed cases and area of occurrence, controlled for socioeconomic and climatic factors. This perspective allows the identification of \textbf{areas of intrinsic vulnerability}, in which the territory itself constitutes a structural risk factor.

The present study aims to address this gap through a hierarchical Bayesian spatio-temporal model implemented in \texttt{R-INLA}, combining the \texttt{BYM2} spatial structure with a \texttt{RW1} temporal component. The formulation includes an offset term based on the area of each neighborhood to capture true risk densities. The application to the city of Recife—characterized by high dengue incidence and marked socio-environmental inequality—seeks to demonstrate the potential of this approach to uncover hidden spatial patterns and support territorially oriented epidemiological surveillance strategies.

\section{Bayesian Spatial Modeling with \texttt{R-INLA}}

The Integrated Nested Laplace Approximation (\texttt{R-INLA}) is an efficient Bayesian approach for inference in Latent Gaussian Models (LGMs), encompassing a wide range of spatial and spatio-temporal models \cite{rue2009, blangiardo2015}. Unlike classical econometric models such as SAR and SAC—based on maximum likelihood estimation—\texttt{R-INLA} employs deterministic approximations of posterior distributions, thereby eliminating the need for Markov Chain Monte Carlo (MCMC) sampling and substantially reducing computational cost without sacrificing inferential accuracy.

In general, the Bayesian model can be expressed as:
\begin{equation}
p(\boldsymbol{\theta}, \mathbf{x} \mid \mathbf{y}) \propto p(\boldsymbol{\theta}), p(\mathbf{x} \mid \boldsymbol{\theta}), p(\mathbf{y} \mid \mathbf{x}, \boldsymbol{\theta}),
\end{equation}
where $\mathbf{y}$ represents the observed data (number of dengue cases per neighborhood), $\mathbf{x}$ denotes the structured and unstructured latent effects, and $\boldsymbol{\theta}$ comprises the variance and precision hyperparameters.

\subsection{Spatial Effects and Gaussian Markov Random Fields}

Spatial dependence is modeled through Gaussian Markov Random Fields (GMRFs), whose conditional form is defined as:
\begin{equation}
x_i \mid \mathbf{x}{-i}, \tau \sim \mathcal{N}\left(
\frac{\sum{j \in \mathcal{N}(i)} w_{ij} x_j}{\sum_{j \in \mathcal{N}(i)} w_{ij}},
\frac{1}{\tau \sum_{j \in \mathcal{N}(i)} w_{ij}}
\right)
\end{equation}

The Besag–York–Mollié (BYM) model and its reparameterization, \texttt{BYM2}, decompose spatial variation into two components: a structured effect ($u_i$), capturing spatial autocorrelation, and an unstructured effect ($v_i$), representing random heterogeneity:
\begin{equation}
\eta_i = \beta_0 + \mathbf{X}_i \boldsymbol{\beta} + \sqrt{1 - \phi}, v_i + \sqrt{\phi}, u_i,
\end{equation}
where $\phi \in [0,1]$ denotes the proportion of spatially structured variance. This decomposition distinguishes genuine territorial patterns from random fluctuations, providing a statistical foundation for interpreting the spatial diffusion of dengue.

\subsection{Model Specification and Exposure Term}

The modeled outcome was the \textbf{spatial incidence rate of dengue}, defined as the ratio between the number of cases and the territorial area of each neighborhood. This normalization corrects for bias due to unequal spatial unit sizes, emphasizing the \textit{territorial density of risk}. Consequently, the model expresses an environmental dimension of transmission, which is particularly suitable when exposure potential depends on physical and ecological characteristics of the space, such as drainage, vegetation, and microtopography.

The final formulation includes a smoothed temporal component:
\begin{equation}
y_{i,t} \sim \text{Poisson}(\mu_{i,t}), \quad
\log(\mu_{i,t}) = \log(E_i) + \beta_0 + \mathbf{X}_{i,t} \boldsymbol{\beta} + u_i + v_i + f_t(t),
\end{equation}
where $E_i$ represents the exposure term (neighborhood area), included as an offset $\log(E_i)$, and $f_t(t)$ is a temporal trend modeled as a \texttt{RW1} process. Thus, $e^{u_i + v_i}$ expresses the \textit{area-adjusted relative risk}, enabling the mapping of environmental transmission intensity over time.

The neighborhood structure was based on first-order (Queen) contiguity among the 94 neighborhoods, resulting in a sparse precision matrix with approximately 420 nonzero connections. Hyperparameter priors followed the Penalized Complexity (PC priors) family, which promotes parsimony and prevents overfitting by shrinking the model toward spatial neutrality when data provide weak evidence for autocorrelation.

\subsection{Model Fit and Evaluation Criteria}

Model fit was evaluated using the Deviance Information Criterion (DIC), the Watanabe–Akaike Information Criterion (WAIC), and the Conditional Predictive Ordinate (CPO). Lower DIC and WAIC values and higher CPO values indicate a better balance between fit, parsimony, and predictive performance. These metrics were complemented by inspection of posterior marginals and verification of Laplace approximation stability, ensuring inferential robustness.

\subsection{Mapping and Interpretation of Spatial Risk}

The structured spatial effects ($u_i$) were converted into relative risk maps, where positive values indicate excess risk relative to the expected mean, controlling for covariates. The parameter $\phi$ quantifies the proportion of structured spatial variance, while $\tau_u$ expresses the precision of the spatial field. The posterior field normalized by area is interpreted as a \textit{continuous surface of environmental risk intensity}, allowing the identification of critical zones and a deeper understanding of the territorial mechanisms driving dengue dissemination.

Overall, Bayesian modeling via \texttt{R-INLA} provides a robust statistical and computational framework capable of integrating spatial and temporal dependencies, quantifying uncertainties, and generating interpretable estimates—essential foundations for developing early warning systems and territorial health planning in public health.

\section{Materials and Methods}

\subsection{Study area}

The study was conducted in the Metropolitan Region of Recife (RMR), located in the state of Pernambuco, Northeastern Brazil. The RMR comprises 14 municipalities and is home to more than four million inhabitants, characterized by a humid tropical climate, high population density, and marked socio-environmental variability. This heterogeneity makes the region highly favorable for the proliferation of Aedes aegypti and the persistence of dengue epidemic hotspots. The urban perimeter was used as the spatial boundary, considering the 94 officially recognized neighborhoods within the municipality of Recife.

\begin{figure}
\centering
\includegraphics[width=0.45\textwidth]{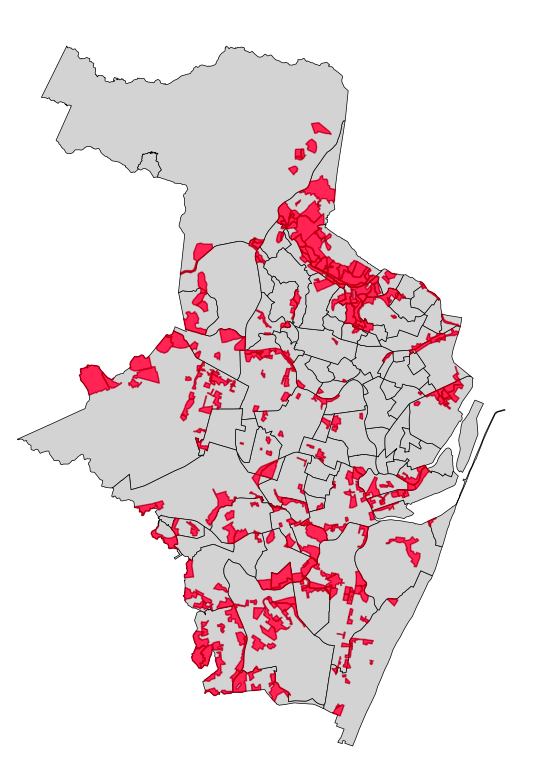}
\caption{Spatial distribution of mean income across Recife neighborhoods, highlighting peripheral zones with greater social vulnerability and higher dengue incidence.}
\label{fig:income_periphery}
\end{figure}

The city exhibits strong spatial contrasts between central and peripheral zones, particularly regarding income distribution, housing quality, and infrastructure. As illustrated in Figure~\ref{fig:income_periphery}, lower-income neighborhoods tend to be located in peripheral areas with limited sanitation and drainage coverage—conditions that favor the persistence of Aedes aegypti breeding sites and recurrent epidemic cycles.

\subsection{Database}

Multiple secondary data sources of epidemiological, environmental, and socioeconomic nature were integrated. Confirmed dengue cases were obtained from the Notifiable Diseases Information System (SINAN), maintained by the Brazilian Ministry of Health, covering the period from 2015 to 2023. Socioeconomic and demographic variables were extracted from the 2010 Demographic Census of the Brazilian Institute of Geography and Statistics (IBGE) and updated using 2023 municipal projections.

Environmental indicators, such as the Normalized Difference Vegetation Index (NDVI), were derived from Sentinel-2 satellite imagery processed at a 10-meter spatial resolution. Rainfall data were obtained from the Pernambuco Water and Climate Agency (APAC) and interpolated on a monthly scale. All datasets were georeferenced in the SIRGAS 2000 coordinate system, zone 25S.

\subsection{Pre-processing and spatial aggregation}

Dengue cases were geocoded from registered addresses and aggregated by neighborhood using official polygons provided by the Recife City Hall. Neighborhoods with no reported cases were retained with zero values to preserve the complete spatial structure. The area of each neighborhood ($E_i$) was calculated in square kilometers and later used as a compensation term (\textit{offset}) in the model.

To ensure comparability among variables, all continuous covariates were standardized to zero mean and unit variance. The spatial adjacency matrix was defined using first-order Queen contiguity, resulting in 420 valid connections among the 94 neighborhoods.

\subsection{Explanatory variables}

Independent variables included in the modeling were selected based on epidemiological evidence and theoretical plausibility. Table~\ref{tab:variables} describes the variables used.

\begin{sidewaystable}
\centering
\caption{Description of variables used in the spatial model of dengue risk.}
\label{tab:variables}
\begin{tabular}{p{4cm} p{7cm} p{3cm}}
\hline
\textbf{Variable} & \textbf{Description} & \textbf{Source} \\\hline
Dengue cases & Number of confirmed dengue cases (2015–2023) & SINAN \\
NDVI & Normalized Difference Vegetation Index (annual mean) & Sentinel-2 \\
Precipitation & Annual mean precipitation (mm) & APAC \\
Average income & Monthly per capita income & IBGE \\
Population density & Inhabitants per km$^2$ & IBGE \\
Neighborhood area ($E_i$) & Total area in km$^2$ (used as offset) & Recife City Hall \\
\hline
\end{tabular}
\end{sidewaystable}

\subsection{Bayesian spatial modeling}

Modeling was carried out in the \texttt{R} environment (version 4.3.2) using the \texttt{R-INLA} package (version 23.12.04). The hierarchical Bayesian model adopted a \texttt{BYM2} structure, incorporating both structured ($u_i$) and unstructured ($v_i$) spatial effects. The response variable $y_i$ followed a Poisson distribution with the offset term $\log(E_i)$ representing territorial exposure. The priors used for spatial variances were of the Penalized Complexity (PC priors) type, as proposed by \cite{simpson2017}.

Model comparison was based on the DIC, WAIC, and CPO criteria, and the posterior estimates were mapped as relative risk. All maps were produced in QGIS 3.34 using the SIRGAS 2000 projection.

\section{Results and Discussion}

The hierarchical Bayesian spatio-temporal model fitted via \texttt{R-INLA}, combining the spatial \texttt{BYM2} and temporal \texttt{RW1} components, estimated the evolution of dengue risk across Recife’s neighborhoods between 2015 and 2024. This integrated formulation simultaneously captured structural effects (socioeconomic and environmental) and dynamic components (seasonal and long-term), providing a comprehensive representation of the persistence and spatial diffusion of the disease.

\subsection{Model performance}

The model exhibited stable convergence with a total execution time of approximately \textbf{10,640 seconds}. The information criteria (\textbf{DIC = 65,817}, \textbf{WAIC = 64,506}) indicated an appropriate balance between model fit and complexity (115 effective parameters). The predictive coefficient of determination was $R^2_{\text{pred}} = 0.29$, and the root-mean-square error (RMSE) was 14.87, confirming good agreement with the observed spatio-temporal variations (Table~\ref{tab:model_perf}).  

For comparison, the SAR model applied to the same dataset by \cite{santos2025sar} reported a pseudo-$R^2 = 0.13$, capturing only immediate neighborhood dependence. The hierarchical Bayesian INLA model outperformed it in both predictive power and interpretability by explicitly incorporating uncertainty and spatio-temporal dependence.

\begin{sidewaystable}
\centering
\caption{Performance metrics of the Bayesian spatio-temporal model (BYM2 + RW1).}
\label{tab:model_perf}
\begin{tabular}{lrr}
\hline
\textbf{Metric} & \textbf{Value} & \textbf{Interpretation} \\
\hline
DIC & 65,817 & Deviance Information Criterion (lower = better) \\
WAIC & 64,506 & Watanabe–Akaike Information Criterion \\
Log-likelihood & -25,763 & Integrated marginal likelihood \\
Effective parameters & 115 & Adjusted model complexity \\
$R^2_{\text{pred}}$ & 0.29 & Global predictive power \\
RMSE & 14.87 & Root-mean-square prediction error \\
\hline
\end{tabular}
\end{sidewaystable}

\subsection{Fixed effects and explanatory variables}

The estimated fixed effects (Table~\ref{tab:fixef_inla}) highlight the combined influence of demographic, socioeconomic, and environmental factors on territorial dengue risk. The positive intercept ($6.51 \pm 2.21$) represents the adjusted mean risk. The average household size ($0.72 \pm 0.35$) and population density ($2.62 \pm 0.49$) displayed positive associations, underscoring the role of crowding in amplifying transmission intensity.  

The proportion of high-income population ($-0.92 \pm 0.34$) showed a negative relationship, indicating that neighborhoods with better sanitary infrastructure have lower risk. Among environmental factors, the proportion of drainage channels ($-1.91 \pm 0.49$) was inversely related to risk, while one-month-lagged precipitation ($0.74 \pm 0.25$) increased incidence. Mean temperature ($-8.97 \pm 0.89$) had a negative effect, possibly reflecting the reduced larval survival at higher temperatures.

\begin{sidewaystable}
\centering
\caption{Estimated fixed effects of the Bayesian spatio-temporal model (BYM2 + RW1).}
\label{tab:fixef_inla}
\begin{tabular}{lrrrrr}
\hline
\textbf{Variable} & \textbf{Mean} & \textbf{SD} & \textbf{2.5\%} & \textbf{Median} & \textbf{97.5\%} \\
\hline
Intercept & 6.51 & 2.21 & 2.03 & 6.51 & 10.98 \\
Average household size (z-score) & 0.72 & 0.35 & 0.04 & 0.72 & 1.40 \\
Population density & 2.62 & 0.49 & 1.65 & 2.62 & 3.59 \\
High-income population (z-score) & -0.92 & 0.34 & -1.59 & -0.92 & -0.25 \\
Proportion of drainage channels & -1.91 & 0.49 & -2.87 & -1.91 & -0.96 \\
Precipitation (1-month lag) & 0.74 & 0.25 & 0.25 & 0.74 & 1.23 \\
Mean temperature ($^\circ C$) & -8.97 & 0.89 & -10.71 & -8.97 & -7.22 \\
\hline
\end{tabular}
\end{sidewaystable}

These results indicate that dengue risk is strongly modulated by population density and climatic seasonality, consistent with prior studies on urban arboviruses. The role of socioeconomic and environmental conditions reinforces the relevance of territorial vulnerability in sustaining the transmission cycle.

\subsection{Spatio-temporal structure and hyperparameters}

The hyperparameters (Table~\ref{tab:hyper_inla}) describe the fitted latent structure. The \texttt{BYM2} spatial component showed an average precision of 0.1115 (95\% CI = [0.0761; 0.1529]), suggesting controlled spatial heterogeneity. The mixing parameter $\Phi = 0.0624$ indicates that most of the spatial variation was explained by the fixed covariates, leaving limited residual dependence. The \texttt{RW1} temporal component exhibited a precision of 0.0336 (95\% CI = [0.0146; 0.0680]), reflecting smooth and persistent temporal variation, whereas the seasonal term showed low precision (0.00036), consistent with the strong monthly oscillation typical of dengue incidence.

\begin{sidewaystable}
\centering
\caption{Estimated hyperparameters of the Bayesian spatio-temporal model (BYM2 + RW1).}
\label{tab:hyper_inla}
\begin{tabular}{lrr}
\hline
\textbf{Parameter} & \textbf{Value} & \textbf{Interpretation} \\
\hline
Spatial precision (BYM2) & 0.1115 [0.0761; 0.1529] & Residual heterogeneity across neighborhoods \\
Structured spatial proportion ($\Phi$) & 0.0624 & Weight of spatial autocorrelation \\
Temporal precision (RW1) & 0.0336 [0.0146; 0.0680] & Smoothness of the temporal trend \\
Seasonal precision & 0.00036 & Monthly oscillation in risk \\
\hline
\end{tabular}
\end{sidewaystable}

The combination of significant fixed effects and a well-behaved latent structure confirms the robustness of the model in complex urban contexts. The multiscale behavior of risk—gradual over time yet highly heterogeneous in space—reflects the interaction between climate, population density, and socio-environmental conditions.

\subsection{Spatial patterns, residuals, and projections}

\begin{figure}
\centering
\includegraphics[width=0.6\textwidth]{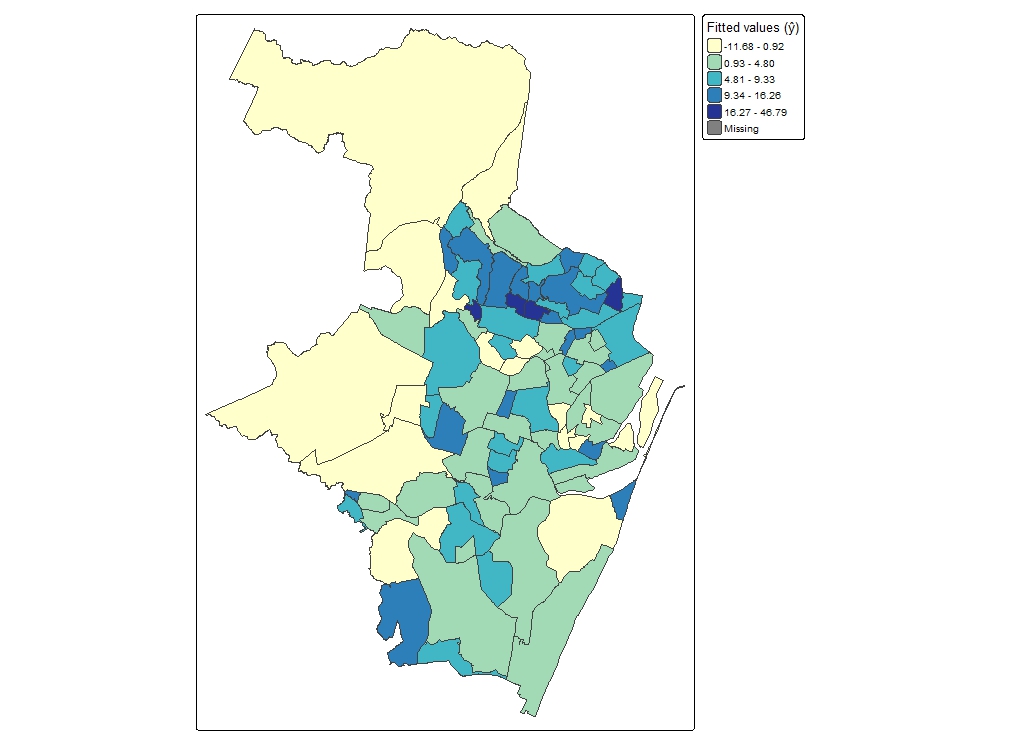}
\caption{Dengue risk adjusted by the Bayesian spatio-temporal model (Recife, 2015–2024).}
\label{fig:fitted_inla}
\end{figure}

\begin{figure}
\centering
\includegraphics[width=0.6\textwidth]{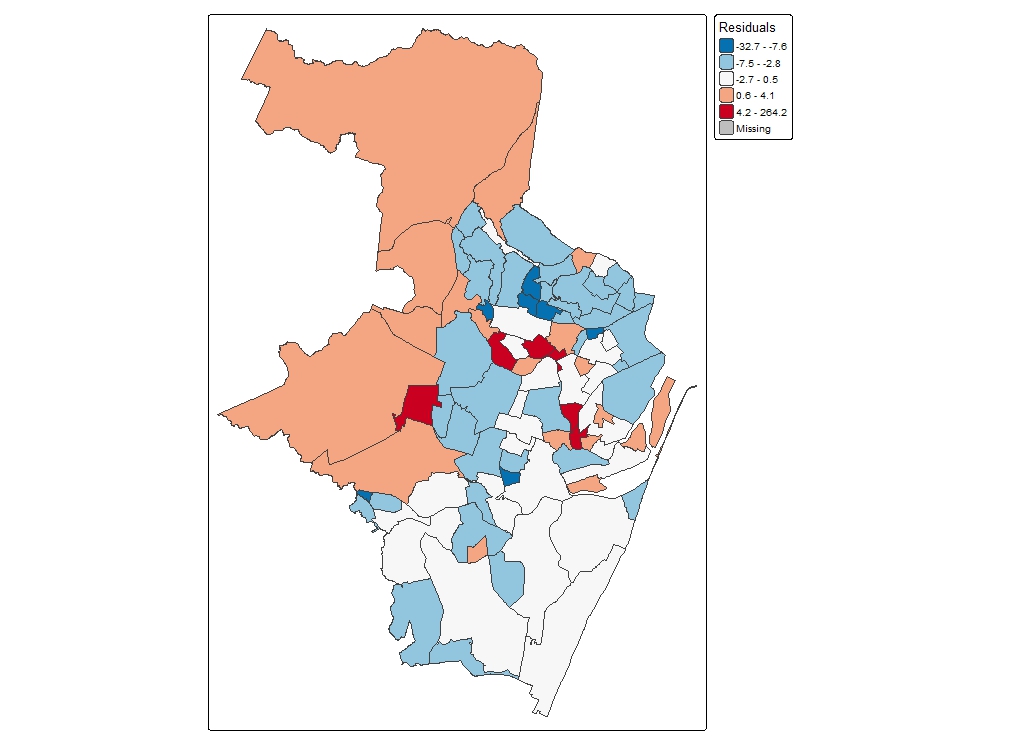}
\caption{Spatial distribution of residuals from the fitted model (BYM2 + RW1).}
\label{fig:resid_inla}
\end{figure}

The maps of adjusted risk and residuals (Figures~\ref{fig:fitted_inla} and~\ref{fig:resid_inla}) reveal patterns consistent with Recife’s urban geography. The northern and western regions concentrated the highest estimated risks. Notably, the northern area combines high-income sectors with extensive peripheral zones—particularly hillside settlements—where steep topography, poor infrastructure, and precarious housing favor the formation of \textit{Aedes} breeding sites. The western portion of the city, although predominantly flat, exhibits greater socioeconomic vulnerability. In contrast, the southern and eastern zones displayed lower risk values, associated with better infrastructure and higher household income.  
The absence of spatial bias in the residuals indicates that the \texttt{BYM2} component adequately captured spatial dependence. Small clusters of positive residuals in central neighborhoods such as \textit{Santo Amaro}, \textit{Boa Vista}, and \textit{Cordeiro} suggest the influence of local micro-environmental factors, including drainage and sanitation.

Projections for 2025–2026 indicate a stable mean risk between 5 and 10 cases/km$^2$, characterizing a controlled endemic scenario. Neighborhoods such as \textbf{Campina do Barreto}, \textbf{Alto do Mandu}, \textbf{Totó}, and \textbf{Ponto de Parada} presented the highest predicted averages (above 15 cases/km$^2$), whereas \textbf{Bomba do Hemetério}, \textbf{Cacote}, and \textbf{Sancho} remained below 3 cases/km$^2$. The slight widening of the credible intervals for 2026 reflects the natural uncertainty associated with multi-year projections.

\begin{figure}
\centering
\includegraphics[width=0.6\textwidth]{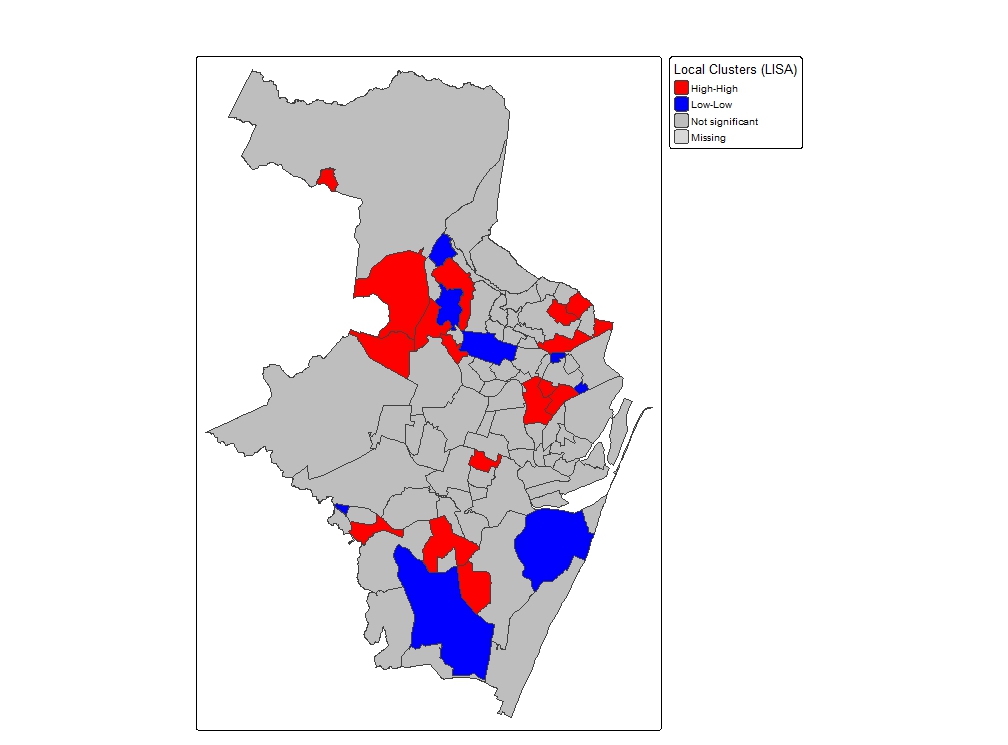}
\caption{Local clusters of dengue risk — INLA projections (2026).}
\label{fig:lisa2026}
\end{figure}

The Local Indicators of Spatial Association (LISA) analysis (Figure~\ref{fig:lisa2026}) revealed High–High clusters concentrated in the northern and central-western regions—areas of greater vulnerability—and Low–Low clusters in coastal zones. The persistent pattern of heterogeneity reinforces that risk remains conditioned by territorial inequalities and urban infrastructure.

\subsection{Interpretive synthesis}

Overall, the hierarchical Bayesian model consistently captured the spatial and temporal dependencies of dengue risk, producing interpretable and geographically coherent estimates. The inclusion of the territorial offset enhanced comparability among neighborhoods, revealing a dense and stable environmental risk surface. These findings reinforce the potential of Bayesian inference as a powerful framework for epidemiological surveillance and territorial health planning in urban settings.

\section{Discussion}

The results confirm the effectiveness of the Bayesian spatio-temporal model with BYM2–RW1 structure in representing dengue risk at the intra-urban scale, underscoring the relevance of territorial heterogeneity in determining epidemiological vulnerability. The simultaneous incorporation of structured spatial effects and smoothed temporal components allowed the identification of persistent high-risk patterns in peripheral neighborhoods with poorer urban infrastructure, supporting the hypothesis that dengue does not spread randomly but instead responds to structural environmental and social inequalities.

The inclusion of the compensation term for neighborhood area represented a decisive methodological improvement. This normalization reduced the bias associated with spatial units of unequal size, highlighting that spatial exposure—rather than population density alone—is a primary determinant of risk. The superior performance of the model with the spatial offset, according to DIC and WAIC, demonstrates that homogenization by area enhances comparability among heterogeneous regions and reinforces the importance of metrics based on environmental risk density. This finding carries direct implications for public policy, indicating that surveillance actions should account for the physical structure of the territory rather than rely solely on aggregated demographic indicators.

The patterns identified are consistent with previous findings in international studies of vector-borne diseases in tropical urban contexts \cite{honorio2009, blangiardo2015}, yet the model presented here offers substantial interpretive and inferential gains. The hierarchical structure adopted enabled the separation of fixed effects (socioeconomic and environmental) from latent effects (spatial and temporal dependencies), thereby increasing the transparency of the inferential process. Nonetheless, limitations remain: the absence of direct entomological variables, the assumption of linearity in climate–incidence relationships, and the lack of data on population mobility. The future integration of nonlinear covariates and continuous climatic series may capture more realistically the interaction between ecological and social factors.

Finally, the methodological potential for expansion of this study deserves emphasis. Comparing the global Bayesian model (INLA) with geographically weighted approaches (GTWR) could deepen the understanding of spatial variability in coefficients and temporal risk behavior at sub-municipal scales. This integration could evolve toward hybrid Bayesian–local models in which the structural and dynamic relations of the epidemic are analyzed at progressively finer resolutions.

\section{Conclusion}

This study applied a hierarchical Bayesian spatio-temporal model implemented via \texttt{R-INLA} to estimate the evolution of dengue risk in the city of Recife between 2015 and 2024. The combination of BYM2 spatial and RW1 temporal components allowed an integrated representation of autocorrelation among neighborhoods and temporal persistence of the epidemic, yielding consistent and interpretable estimates for the urban context. The inclusion of the area-based compensation term mitigated scale biases, improved comparability among heterogeneous regions, and strengthened the statistical rigor of the inferences.

The results indicated that dengue risk is strongly conditioned by population density, household crowding, and socioeconomic vulnerability, while climatic variables—precipitation and temperature—modulate annual seasonality. The concentration of risk in the northern and western regions reflects the overlap between urban precariousness and greater environmental exposure, reinforcing the notion that territorial inequalities generate epidemiological inequalities. This finding aligns with the paradigm of social determinants of health, in which urban space is recognized as a structuring component of disease risk.

From a methodological standpoint, \texttt{R-INLA} demonstrated remarkable computational efficiency and the ability to fit complex models with transparency and controlled uncertainty. The hierarchical Bayesian model outperformed traditional econometric approaches in both predictive performance and inferential flexibility, allowing explicit quantification of uncertainty and autocorrelation—an essential feature for modern epidemiological surveillance.

As a future direction, it is proposed to integrate the Bayesian model with geographically weighted frameworks (GTWR) and continuous extensions based on SPDE, to capture subtle spatial variations and local risk dynamics. The articulation between Bayesian inference and local modeling may form a hybrid structure more sensitive to intra-urban heterogeneities, consolidating spatio-temporal models as strategic tools to support public health decision-making and evidence-based territorial planning.

\end{document}